# A skeletonization based image segmentation algorithm to isolate slender regions in 3D microstructures

Vinit Vijay Deshpande[1] and Romana Piat[1]

[1] Department of Mathematics and Natural Sciences, University of Applied Sciences Darmstadt, Schöfferstraße 3, Darmstadt 64295, Germany

**Abstract:** The work proposes an image segmentation algorithm that isolates slender regions in three-dimensional microstructures. Characterizing slender regions in material microstructures is an extremely important aspect in material science because these regions govern the macroscopic behavior of materials for many applications like energy absorption, activation of metamaterials, stability of high temperature filters, etc. This work utilizes skeletonization method to calculate centerline of the microstructure geometry followed by a novel pruning strategy based on cross-sectional area to identify slender regions in the microstructure. 3D images of such microstructures obtained from micro-CT often suffer from low image resolution resulting in high surface noise. The skeleton of such an image has many spurious skeletal branches that do not represent the actual microstructure geometry. The proposed pruning method of cross-sectional area is insensitive to surface noise and hence is a reliable method of identifying skeletal branches that represent the slender regions in the microstructure. The proposed algorithm is implemented on a test case to showcase its effectiveness. Further it is implemented on a 3D microstructure of ceramic foam to identify the slender regions present in it. It is shown that the method can be used to segment slender regions of varying dimensions and to study their geometric properties.

## 1. Introduction

With the advent of micro computed tomography (CT), microstructures of a variety of heterogeneous materials can be obtained in the form of three-dimensional (3D) images. If the different phases (or materials) in the heterogeneous medium have good contrast between them, they can be segmented easily. However, identifying slender regions within the same phase is very challenging because they are connected to other non-slender (or thick) regions of the same phase. These slender regions also have varying dimensions, so the segmentation procedure has to be independent of any fixed kernel size. Examples of such materials are shown in Fig.1. In these examples, the slender regions play a critical role in the material performance. In mechanical applications, these slender regions are often the weakest members and hence characterizing them is critical in predicting material performance. In applications like metamaterials, these slender regions determine the activation regime. However, often they are interconnected to other slender regions or thick regions. Hence, isolating them becomes an important precursor to microstructure characterization.

Isolating slender regions from a 3D image has been a widely studied in composite materials. In the case of fiber reinforced composites, the orientation and length distribution of fibers in a matrix can be determined by applying techniques like anisotropic Gaussian filtering, Hessian matrix calculation and structure tensor calculation on computed tomography data of the composite [1]. All these techniques make use of a kernel filter whose shape and size is dependent on the shape and size of the fiber which is known in advance and is constant in the microstructure. The effectiveness of such techniques drastically reduces if the slender regions in the microstructure have a wide distribution in their shape and sizes.

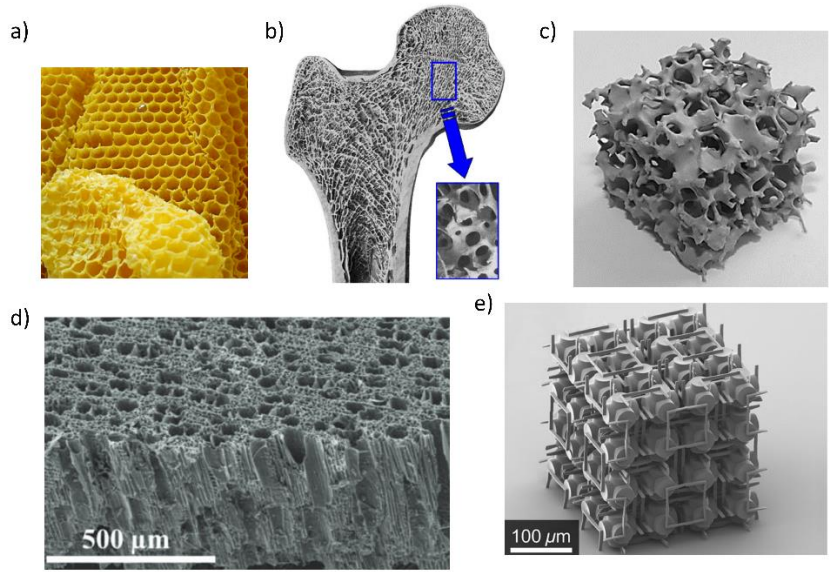

**Fig.1.** a) *Honeycomb structure; microstructure of b) human femur bone [2]; c) ceramic foam derived from potassium based geopolymer [3]; d) carbonized wood [4] and e) metamaterial [5].*

Classic image segmentation methods rely on grayscale values of images to perform segmentation [6]. Methods based on edge detection utilize the differential in the gray scale value to find the edges of objects [7]. Other methods based on regions division like thresholding [8, 9, 10], region growing [11, 12] utilize statistics of grayscale values in the image like standard deviation, variance, etc. to segment image features. However, segmenting specific features within the same phase or same object in the image that does not have any difference in grayscale values is still a challenge.

Skeletonization algorithms [13] convert a 3D binary image into a skeleton image in which the skeleton represents the medial axis of the image geometry. There are different definitions of skeletonization in literature. [14] defined skeleton as a set of voxel points at which an advancing firefront gets extinguished by itself. [15] defined skeleton as a set of centers of balls with maximum radius that can lie entirely within the domain of the studied structure. In general, a skeleton should be topologically equivalent to the original structure, it should have 1 voxel thickness and it should be centrally located (medially) to the structure [16]. Skeletons are generally used in applications like pattern recognition [17], studying tortuosity of connected structures [18], medical image segmentation [19], etc.

Widely used skeletonization algorithms are based on distance transform [20] which gives minimum distance of each voxel in the domain of the studied region (foreground) from the background. When the result of distance transform is superimposed on the skeleton of the image, it is used to study the thickness and topology of the image geometry. This information can potentially be used to isolate slender regions in the image on the basis of the calculated thickness. However, the distance transform is highly sensitive to image noise and insufficient resolution. In many applications, the resolution of micro-CT is not sufficient enough to capture the detailed features of the material microstructures. Hence, direct application of skeletonization leads to a very hairy skeleton which has many spurious branches that do not represent the actual geometry of the structure. To overcome this, regularization of the skeleton is performed by different pruning strategies. Pruning strategies can be divided into two main categories. One which modifies the boundary surface of the structure [21,22,23,24] to reduce the noise and the other which remove unwanted skeletal branches on the basis of some significance criteria assigned to each skeletal branch [25,26,27,28,29]. The first category which basically smoothens the boundary surface can change the skeleton drastically by shifting its position and/or by creating new spurious branches. This smoothing can also alter the boundary surface significantly leading to change in the topology of the structure. The second category is based on assigning specific values called as significance measures to each skeletal branch. These values are generally geometric quantities that define the amount of change it will cause to the topology of the structure if that skeletal branch is pruned. Examples are propagation velocity of the symmetry axis, erosion thickness, erosion area, etc. These significance measures are always application dependent and hence cannot be generalized as effective strategies.

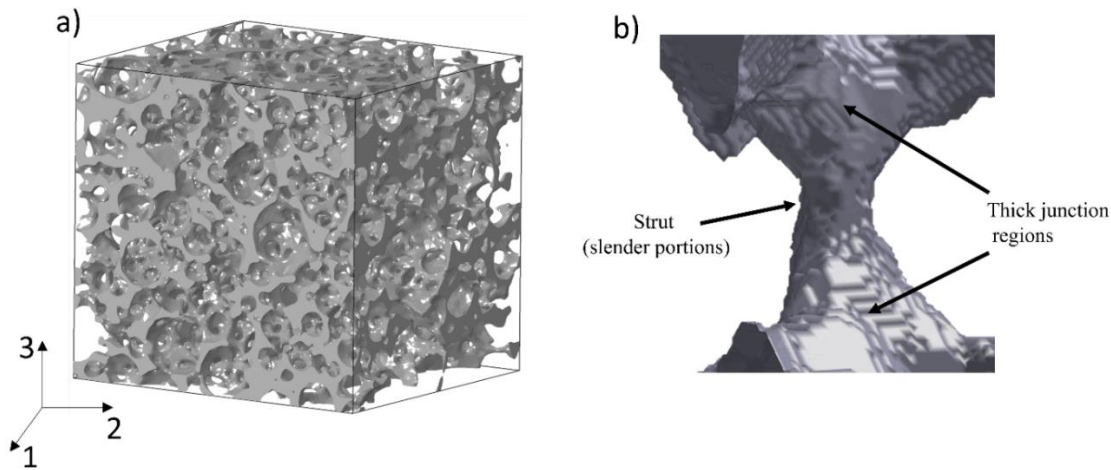

**Fig.2.** *a) A cubic sample of the microstructure; b) a slender portion called strut connected to thick junction regions.*

In this work, the skeletonization algorithm is utilized as a first step to isolate slender regions in a 3D microstructure image. A novel pruning strategy based on cross-sectional area of the region is devised that identifies slender regions not on the basis of their thickness but on the basis of their cross-sectional area. It will be proved that this pruning strategy is insensitive to image noise and hence is a robust method to isolate the slender regions. The example of material microstructure studied in this work is ceramic foam. The microstructure of ceramic foam in the form of an image was obtained by micro-CT scanning a material sample [30]. The details regarding binarization and image characterization of this sample can be found in [31]. The slender regions in this microstructure which are also known as struts play a critical role in the mechanical performance of this material [32]. A small cubic sample of this foam microstructure is shown in Fig.2a. The volume fraction of the ceramic in the foam is 0.255. It can be seen that the microstructure is made up of slender regions called struts which are connected to thick junction regions. An example of such a strut is shown in Fig.2b. The objective of the developed algorithm is to segment these struts in the microstructure.

This paper is organized as follows: section 2 describes the basic skeletonization algorithm, novel pruning strategies and an algorithm to determine volume of the slender portions, section 3 shows the effectiveness of these strategies on a simple test case, section 4 describes the results of implementation of algorithm on the ceramic foam material microstructure and section 5 concludes the article.

## 2. Skeletonization and pruning strategies

### 2.1 Skeletonization

[8] proposed a distance ordered homotopic thinning method for skeletonization that maintained the topology of the structure throughout the thinning process. Fig.3a shows a 2D binary image for illustration purpose. Let S and $\bar{S}$ be a set of voxels with value 1 and 0 respectively. Let p be any point (voxel) in the 3D image such that $p = (x, y, z) \in \mathbb{Z}^3$.

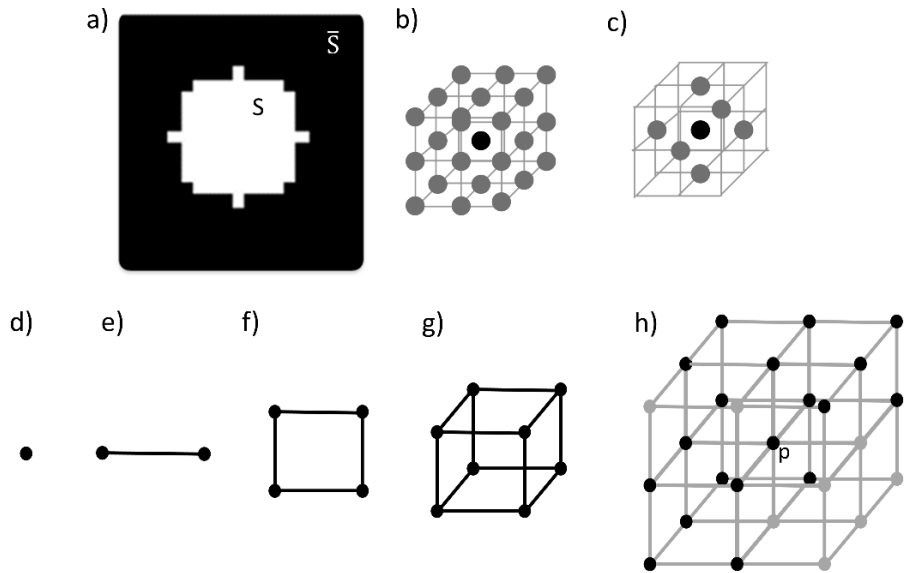

**Fig. 3.** *a) 2D binary image for illustration; b) 26- and c) 6-neighbourhood (grey spheres) of a point (black sphere); a) a single point called vertex v; b) an edge e formed by 2 points; c) a face f formed by 4 points; d) an octant oct formed by eight points and e) 26-neighbourhood of point p.*

In order to understand the skeletonization algorithm, certain definitions are introduced as followed. First is the neighborhood of a point p which is defined as,

$$N_{26}(p) := \{p' \mid \max(|x - x'|, |y - y'|, |z - z'|) \leq 1\} \quad (1)$$

$$N_6(p) := \{p' \mid (|x - x'| + |y - y'| + |z - z'|) \leq 1\} \quad (2)$$

Fig. 3b shows 26 neighbors (grey spheres) to the point p (black sphere) and Fig. 3c shows 6 neighbors (grey spheres) to the point p (black sphere). Next, points that are inside each other's n-neighborhood are called n-adjacent points. Two points p and p' are said to be n-connected if there is a sequence of points $p_0(=p), p_1, \ldots p_k(=p')$ such that each $p_i$ is n-adjacent to $p_{i-1}$ for $1 \leq i \leq k$. Based on these definitions of connectedness, an object O is defined as a set of n-connected points in S. A cavity C is defined as an object in $\bar{S}$ that is completely surrounded by points in S. A hole H is defined as a tunnel (open on both sides) through S. In following text, the sign "#" corresponds to "cardinality of a set". By calculating the number of objects #O, number of holes #H and number of cavities #C in a 6-connected set of points, the Euler characteristics $G_6$ of these points is defined as [33]:

$$G_6 := \#O - \#H + \#C. \quad (3)$$

It is also possible to define $G_6$ in terms of simplices. Examples of simplices are a single point called vertex $v$, an edge $e$ formed by 2 points, a face $f$ formed by 4 points and an octant $oct$ formed by 8 points. These simplices are shown in Fig.3d-g. By calculating the number of vertices $\#v$, number of edges $\#e$ and number of faces $\#f$ and number of octants $\#oct$ in a 6-connected set of points, the Euler characteristics $G_6$ of these points is defined as [33]:

$$G_6 = \#v - \#e + \#f - \#oct. \quad (4)$$

If we only consider the case in which objects in S are 26-connected and those in $\bar{S}$ are 6-connected, the Euler characteristic of 26-connected S is be related to that of a 6-connected $\bar{S}$ by,

$$G_{26}(S) = G_6(\bar{S}) - 1. \quad (5)$$

Next, a simple point is defined as that point in S which if deleted, i.e. changed from voxel value 1 to 0, does not change the topology of S. It was shown in [34] that any border point p in S is called a simple point if its deletion does not change the number of objects and holes in S and $\bar{S}$, i.e.

$$\delta O(S) = 0, \delta O(\bar{S}) = 0, \delta H(S) = 0, \delta H(\bar{S}) = 0 \Leftrightarrow \text{p is simple}, \quad (6)$$

where $\delta$ denotes change in the property after deletion of the simple point. It was proved in [25] that these conditions on simple point are equivalent to not changing the number of objects and Euler characteristics in the 26-neighbourhood of point p in S, i.e.

$$\delta O(S \cap N_{26}(p)) = 0, \delta G_{26}(S \cap N_{26}(p)) = 0 \Leftrightarrow \text{p is simple}. \quad (7)$$

Fig.3h shows a 26-neighbourhood of point p such that the black points belong to S and the grey points belong to $\bar{S}$. This way the entire points in set S are studied to identify simple points and to delete them. This process continues till there no more simple points present in the set S. It can be seen that in order to identify a simple point, only its 26-neighbourhood is studied. This allows parallelization of the process and yields much faster results.

## 2.2 Pruning strategies

Since the skeletonization algorithm is sensitive to noise in the image data, different pruning strategies are required to remove the unwanted spurious skeletal branches. The different pruning strategies developed in this work are described in Fig. 4. The first strategy is pruning free branches. Any branch of the skeleton that is connected to the main body of skeleton at only point is defined as a free branch. The next pruning strategy is based on significance measures [27]. They are geometric measures namely length of the branch, curvature of the branch, minimum cross-sectional area, eccentricity and aspect ratio of the strut (slender portion) that the branch represents.

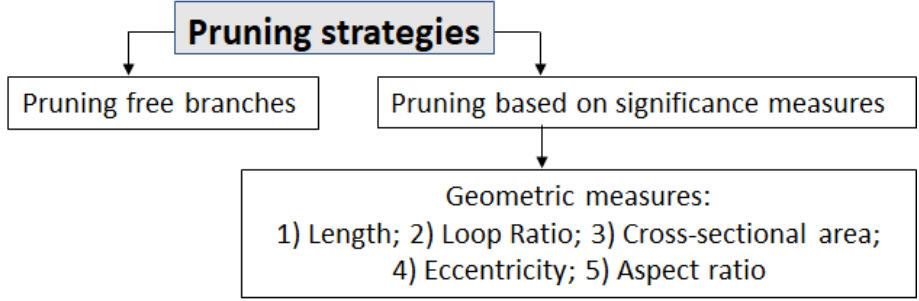

**Fig.4.** *Pruning strategies implemented in this work.*

## 2.3 Pruning free branches

The peculiar property of the microstructure shown in Fig. 2a is that all the struts are connected to thick junction regions at both the ends (refer Fig.2b). However, when the material sample is cut into a cubic shape, the struts lying at the boundary also get cut and therefore have one free end that touches the boundary. Hence, the skeletal branch of such a boundary strut also has one free end. The skeletal branch that represents a strut lying entirely in the interior of the cubic domain is always attached to other skeletal branches at both ends. However, there are many other skeletal branches in the interior of the cubic domain that have one free end. These are spurious branches that do not represent the actual structure. In this section, the objective is to remove these spurious skeletal branches.

Consider the Fig.6a. Let Q be a set of points belonging to the skeleton of the studied structure and $N_{26}(q)$ is the neighborhood of q.

$$Q := \{q = (x, y, z) | q \in S \text{ at the end of skeletonization algorithm}\} \quad (8)$$

Let $W = \{w\}$ be a set of junction points where the different skeletal branches meet (see Fig. 5a).

$$\#v\big(Q \cap N_{26}(q)\big) \geq 3 \implies w = \text{junction point} \quad (9)$$

It means all the points in Q that have at least 3 neighbors in its 26-neighbourhood are defined as junction points. Let $\mathbb{T}$ be a set of objects formed by the skeletal branches such that $\mathbb{T} = O(Q \setminus W)$.

$$\mathbb{T} := \{T(i) \in Q \setminus W \mid 1 \leq i \leq n_s\}, \quad (10)$$

where $n_s$ is the total number of skeletal branches. Let $T_f \subseteq \mathbb{T}$ be defined as a free branch such that,

$$\#(q \in T(i) \mid \#(N_{26}(q) \cap W) \neq 0) = 1 \implies T_f \text{ is a free branch.} \tag{11}$$

It means that $T_f$ is a free branch if it is connected to only one junction point (refer Fig.6b). However, we do not want to prune the free branches that touch the boundary of the cubic domain (refer Fig.6c). Therefore, only free branches that lie entirely in the interior of the cubic domain are pruned. Hence,

$$\{\forall q \in T_f \mid (|x_q - x_{max}| > 0 \,\&\, |x_q - x_{min}| > 0 \,\&\, |y_q - y_{min}| > 0 \,\&\, |y_q - y_{max}| > 0 \,\&\, |z_q - z_{min}| > 0 \,\&\, |z_q - z_{max}| > 0 )\} \implies \text{a free branch } T_f \text{ is pruned (removed)}, \tag{12}$$

where, $(x_q, y_q, z_q)$ are the coordinates of point q and $(x_{min}, y_{min}, z_{min})$ and $(x_{max}, y_{max}, z_{max})$ are the diagonally opposite corners of the cubic domain where the first corner is closest to the origin and the second corner is the farthest.

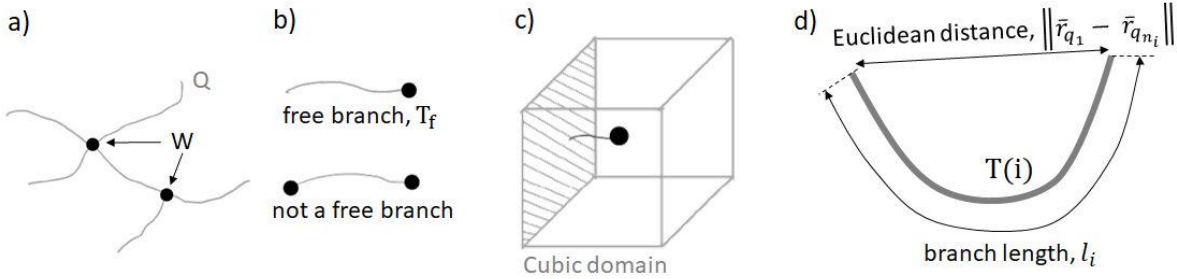

**Fig.5.** *a) A skeleton set Q and junction point set W; b) difference between free branch and not free branch; c) a free branch that touches the boundary of the cubic domain; d) illustration to measure loop ratio of an example skeletal branch.*

.

## 2.4 Pruning based on geometric significance measures

### 2.4.1 Length of the skeletal branch

The length of each $i^{th}$ skeletal branch $T(i)$ in the digital image is defined as number of voxels $\#(T(i))$ present in it. Therefore,

$$l_i := \#(T(i)). \tag{13}$$

Any skeletal branch is pruned if its length is less than a threshold value $\lambda_{length}$, i.e.

$$l_i \leq \lambda_{length} \implies \text{the skeletal branch } T(i) \text{ is pruned.} \tag{14}$$

### 2.4.2 Loop ratio of the skeletal branch

Some skeletal branches in the microstructure have such a high curvature that they so not represent an actual strut in the microstructure. In order to quantify this, a parameter called Loop Ratio (LR) is defined. It is the ratio of Euclidean distance between the end points $q_1$ and $q_{n_i}$ of the skeletal branch T(i) to the length $l_i$ of the skeletal branch (refer Fig. 5d). Then for the Loop Ratio of $i^{th}$ skeletal branch:

$$\text{LR}(i) := \frac{\|\bar{r}_{q_1} - \bar{r}_{q_{n_i}}\|}{l_i} \text{ where } q_j \in T(i) \mid 1 \leq j \leq n_i \ , \tag{15}$$

$\bar{r}_q$ is the position vector of point q and $n_i$ is the total number of points in skeletal branch T(i).

The pruning criterion is defined as,

$$\text{LR}(i) \leq \lambda_{LR} \Rightarrow \text{the skeletal branch T(i) is pruned.} \tag{16}$$

### 2.4.3 Cross-sectional area of the strut

Fig. 6a demonstrates the calculation of the cross-sectional area of an example strut $i$. q is any point on the skeletal branch T(i), i.e. $T(i) = \{q_j \mid 1 \leq j \leq n_i\}$. $\bar{r}_q$ is the position vector of point q and $\bar{n}_j = \bar{r}_{q_j} - \bar{r}_{q_{j-1}}$ is the tangent vector to the skeleton at point $q_j$. Let $\mathbb{B}(\bar{n}, q)$ be a set of all the points in S that lie in the cross-sectional plane normal to $\bar{n}$ and going through point q. These set of points are determined by the following condition.

$$\mathbb{B}(\bar{n}, q) := \{p \in S \mid \bar{n} \cdot (\bar{r}_p - \bar{r}_q) = 0\}. \tag{17}$$

This cross-sectional plane is shown diagrammatically in Fig.6a and an example of such a plane of the studied microstructure is shown in Fig.6b.

Let $\mathbf{C}_\mathbb{B} = O(\mathbb{B})$ be a set of 26-connected objects in $\mathbb{B}$. i.e. $\mathbf{C}_\mathbb{B} = \{C_\mathbb{B}(i) \mid 1 \leq i \leq m\}$ where $m$ is the total number of connected objects in $\mathbb{B}$. The next task is to find that object $C_\mathbb{B}^q$ in $\mathbb{B}$ to which any point q belongs to i.e.

$$q \subseteq C_\mathbb{B}(i) \Rightarrow C_\mathbb{B}^q = C_\mathbb{B}(i). \tag{18}$$

Cross-sectional area of the strut T(i) at point $q_j$ (refer Fig.46b) is given as,

$$A_j = \#\left(C_\mathbb{B}^{q_j}\right). \tag{19}$$

Lastly, minimum cross-sectional area of the strut $A_{min}(i)$ is defined as,

$$A_{min}(i) = \min_{1 \leq j \leq n_i} (A_j). \tag{20}$$

The pruning criterion is defined as,

$$A_{min}(i) > \lambda_{minA} \Rightarrow \text{the skeletal branch is pruned.} \tag{21}$$

After pruning, in all the remaining skeletal branches, it can be seen that since the branch extend from junction point to junction point, it represents not only the thin region of the strut but also the thick junctions of the material microstructure. An example of such strut is shown in Fig. 6c. Since the objective of this study is to segment the slender portions, it is important to define which part of the material structure that the skeletal branch represents is an actual strut and which part is a material junction. It is done by selecting only that part of the strut whose cross-sectional area satisfies the below condition.

$$A_j < \lambda_s \cdot A_{\min}(i). \tag{22}$$

In Fig.6c, the skeletal branch is shown by grey color markers that connect the two junction points. A small part of this skeletal branch is selected based on its cross-sectional area as shown in Fig. 6d and Eq.22. This small part of the strut is highlighted by black color markers in Fig. 6c.

Hence, the part of the skeletal branch that represents this slender portion is defined as,

$$T^{\text{new}}(i) := \{q_j \in T(i) \mid A_j < \lambda_s \cdot A_{\min}(i)\}. \tag{23}$$

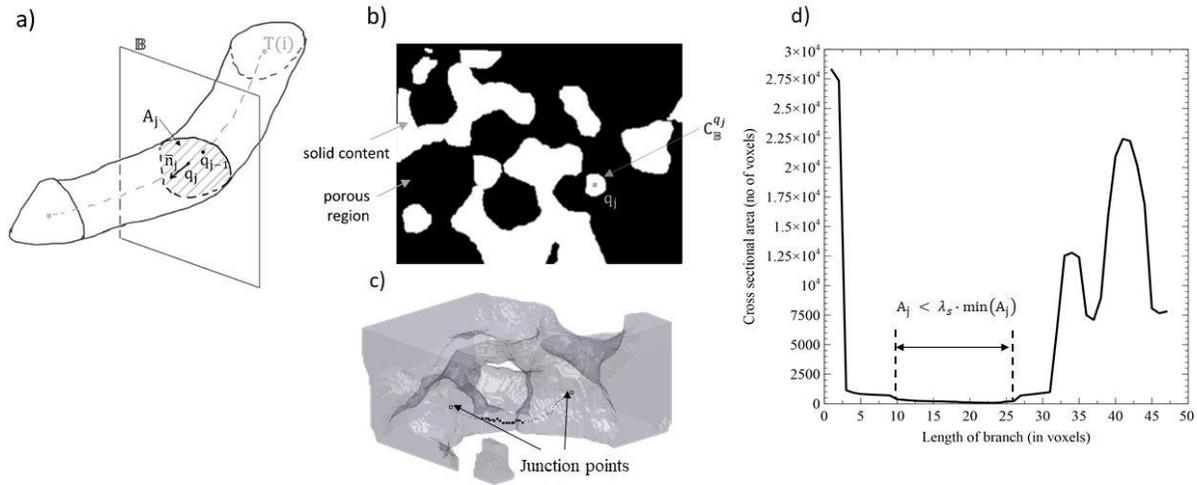

**Fig.6.** *a) Description of cross-sectional area of an example strut; b) plane $\mathbb{B}$ showing cross-section of the microstructure; c) an example of a strut in the studied microstructure and d) cross-sectional area of the strut along its length.*

### 2.4.4 Eccentricity of the cross-section of the strut

It is observed that in the microstructure region where thick junctions are converted into skeletons, sometimes the skeletons do not represent the actual medial axis but are actually offset to the cross-section. This is again because of the non-smooth nature of the material surface. In order to identify such skeletal branches, a parameter called equivalent radius is defined. At each point $q_j$ of the new branch $T^{\text{new}}(i)$, an equivalent radius, $R_j^{\text{eq}}$ is calculated as,

$$R_j^{\text{eq}} = \sqrt{\frac{A_j}{\pi}}. \tag{24}$$

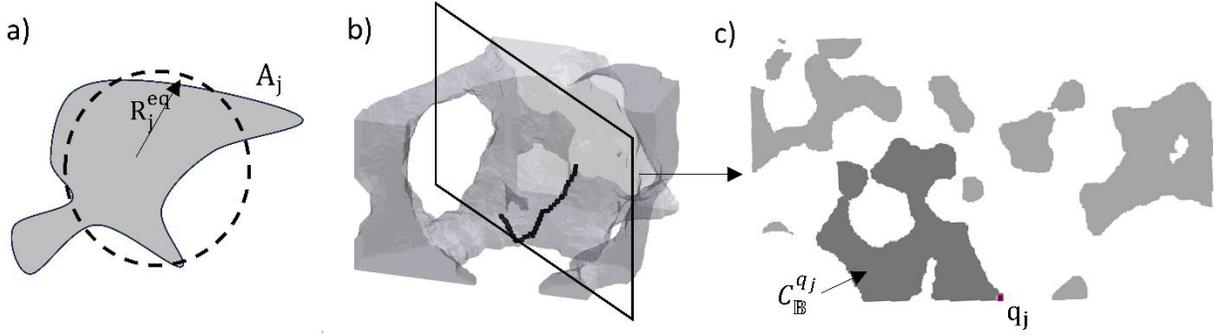

**Fig.7.** *Relationship between equivalent radius, $R_j^{eq}$ and cross-sectional area, $A_j$ ; b) an example of a skeletal branch with very high eccentricity; c) cross-section at a particular skeletal point.*

Fig. 7a illustrates the relationship between $R_j^{eq}$ and $A_j$. Let the thickness of the strut at each point $q_j$ be defined as $t_j$ which is calculated from the distance transform of the image [12]. The eccentricity at each point $q_j$ is defined as,

$$\text{Eccen}_j = \frac{R_j^{eq}}{t_j}. \qquad (25)$$

Note that in Eq.25, division is carried out elementwise. Higher the value of this parameter, the more offset will be the position of point $q_j$ with respect to the centre of the cross-sectional area. An idealized cylindrical strut will have eccentricity of 1 along its entire length. An example of a skeletal branch in the microstructure is shown in Fig. 7b. The cross-section of the strut (object $C_\mathbb{B}^{q_j}$) corresponding to the highest value of $\text{Eccen}_j$ is shown in dark grey color in Fig. 7c. It can be seen that the object $C_\mathbb{B}^{q_j}$ does not belong to any particular strut but to a thick junction. The point $q_j$ certainly does not represent the medial point of this object. Such skeletal branches are pruned using the parameter of maximum eccentricity.

Maximum eccentricity of any strut $T^{new}(i)$ is given as,

$$\text{Eccen}_{max}(i) := \max(\text{Eccen}_j). \qquad (26)$$

The pruning criterion is defined as,

$$\text{Eccen}_{max}(i) \geq \lambda_{eccen} \Rightarrow \text{the skeletal branch is pruned.} \qquad (27)$$

### 2.4.5 Aspect ratio of the strut

It is observed that some skeletal branches having very low aspect ratios do not represent the real struts but are artefacts due to limited resolution of the digital image. The aspect ratio $AR(i)$ of any strut $T^{new}(i)$ is defined as,

$$AR(i) = \frac{\#\left(T^{new}(i)\right)}{2\min(t_j)}, \tag{28}$$

where $t_j$ is thickness of the strut at point $q_j$. A typical strut in shown in Fig. 8a. The thickness for the aspect ratio calculation is taken as the minimum thickness of the strut. Skeletal branched with very low aspect ratio are pruned according to the following condition.

$$AR(i) \leq \lambda_{AR} \Rightarrow \text{the skeletal branch is pruned.} \tag{29}$$

## 2.5 Determination of strut volume

The algorithm till now identifies all the skeletal branches that are not pruned. The next step is to identify the strut volume in the digital image that corresponds to each branch. Let p be any point in S. A spherical neighborhood of point q (refer Fig. 8a) is defined as,

$$N^{sph}(q) := \{p \in S \mid \|\bar{r}_p - \bar{r}_q\| \leq t\}, \tag{30}$$

where t is the thickness of the strut at point q. Let $TR(i)$ be the set of strut voxels that correspond to $T^{new}(i)$. A new set $TR_a(i)$ is defined by selecting all the points (voxels) in S that lie in the spherical neighbourhood of all the points in the skeletal branch $T^{new}(i)$ (refer Eq. 23).

$$TR_a(i) := \{p \in N^{sph}(q_j) \mid 1 \leq j \leq \#\left(T^{new}(i)\right)\}. \tag{31}$$

This set contains all the strut voxels in the interior of the strut. A 2D illustration of this set with an example strut in shown in Fig. 8b. However, at the end points of the skeletal branch $T^{new}(i)$, it contains some extra voxels as a result of the definition of $N^{sph}(q)$. To remove these extra voxels, another set $TR_b(i)$ in defined that contains all the voxels in S that lie in between the cross-sectional planes of the end points (refer Fig. 8a) of the skeletal branch $T^{new}(i)$. This set is defined as,

$$TR_b(i) := \{p \in S \mid \bar{n}_1 \cdot (\bar{r}_p - \bar{r}_{q_1}) > 0 \ \& \ \bar{n}_{end} \cdot (\bar{r}_p - \bar{r}_{q_{end}}) < 0\}. \tag{32}$$

The subscripts '1' and 'end' mean the first and the last point of skeletal branch $T^{new}(i)$. A 2D illustration of this set with the same example strut in shown in Fig. 8c. Finally, the intersection of these two sets (refer Fig. 8d) gives a set that contains all the strut voxels $TR(i)$ that correspond to skeletal branch $T^{new}(i)$, i.e.

$$TR(i) = TR_a(i) \cap TR_b(i) \tag{33}$$

An example of a strut from the microstructure is shown in Fig. 8e. The light grey color regions are the foam microstructure. The dark grey color represents strut set TR(i) and the back color voxels represent set $T^{new}(i)$.

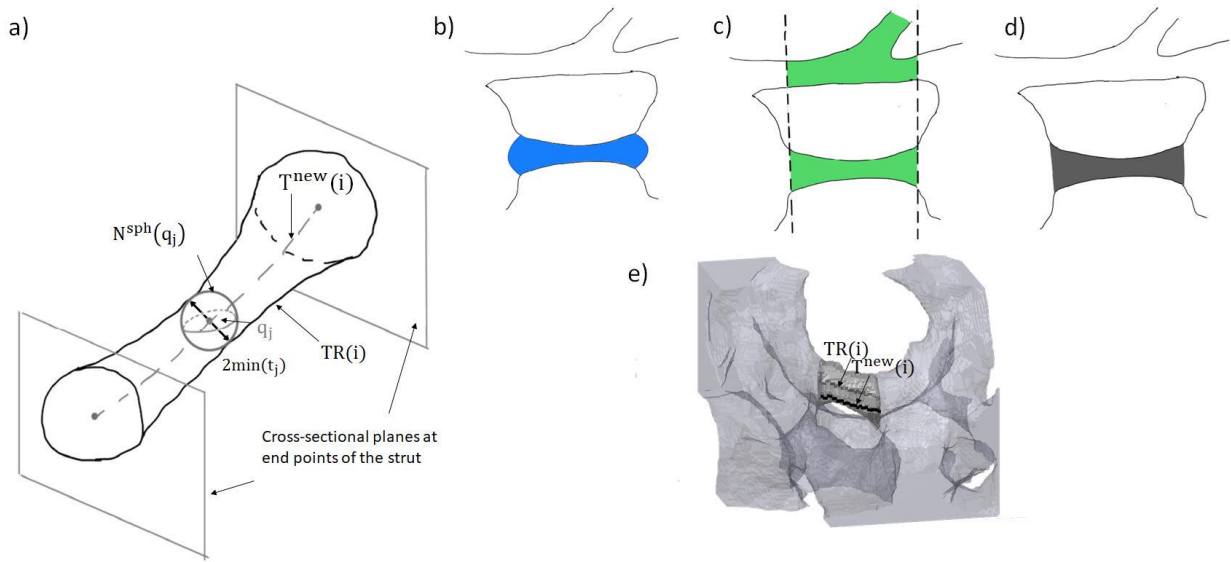

**Fig.8.** *a) Illustration of aspect ratio of an example strut, a typical strut describing $TR_a(i)$ and $TR_b(i)$; 2D illustration of set b) $TR_a(i)$ , c) $TR_b(i)$ and d) $TR(i)$ ; e) volume region of $TR(i)$ of an actual strut in the microstructure.*

## 3. Numerical results: Implementation on a test case

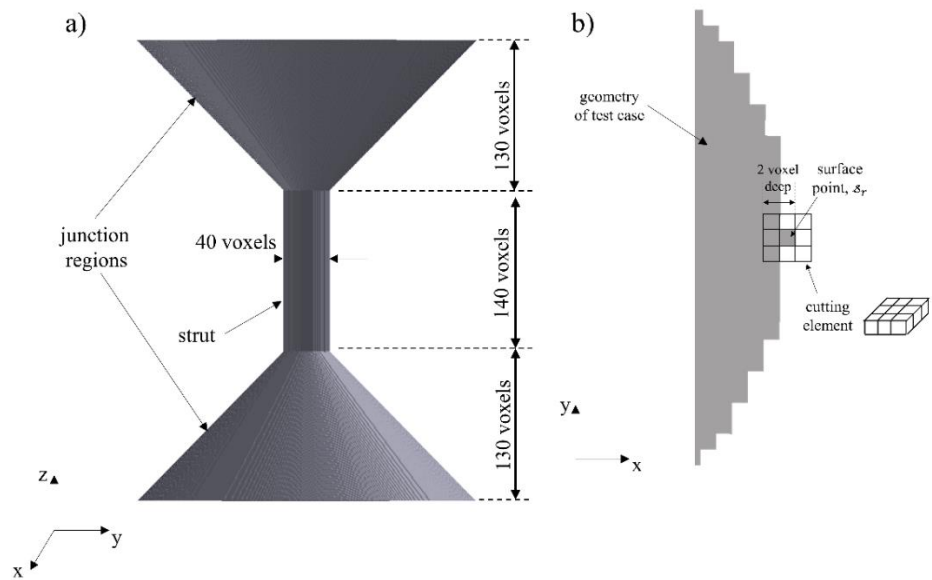

**Fig.9.** *a) A simplified geometry resembling a strut and the junction regions; b) definition of a cutting element.*

The studied 3D microstructure of ceramic foam is very complicated to visualize. It essentially consists of slender regions that connect to thick junction regions at both ends. To simulate this, a simple test case is defined in which a cylindrical strut (slender region) has conical junction regions at both ends as shown in Fig. 9. The strut has a circular cross-section with diameter chosen as 40 voxels for demonstration purposes. The performance of the skeletonization algorithm and the pruning strategies are evaluated on this standard test case.

The main disadvantage of skeletonization algorithm is its susceptibility to noise on the surface caused by lack of sufficient image resolution. With the help of this test case, it will be shown how the surface noise affects the results of the skeletonization algorithm and how the pruning strategies rectify this issue and identify the strut region correctly.

The noise of the geometry surface is artificially introduced using the following procedure. All the voxel points on the surface are identified as points that have at least one 26-neighbour belonging to the other phase (or background) $\bar{S}$. This set is defined as $\mathcal{S}$, such that

$$\mathcal{S} := \{s \in S \mid N_{26}(s)) \cap \bar{S} \neq \emptyset\} \tag{34}$$

A small subset of these points, $\mathcal{S}_r \subseteq \mathcal{S}$ is randomly selected and a cutting element is defined at each of these points $s_r$ as shown in Fig. 9b. The figure shows a cross-section of the strut (in gray color) and the location of the cutting element on the surface point. It is a square shaped cutting element with 1 voxel thickness along z-axis and square shape in x-y plane. This cutting element deletes any point of the geometry (assigns voxel value 0) that it intersects with. The noise of the surface is increased by increasing the edge length of the square. 2% of the total surface points are randomly selected to define the cutting elements, i.e. $\#(\mathcal{S}_r) = 0.02\#(\mathcal{S})$. Three edge lengths of the cutting element are considered in this study, namely, 3 voxels, 5 voxels and 7 voxels. The central voxel of the cutting element is always located on the surface voxel, $s_r$ of the geometry. This way, the larger the size of the cutting element, the deeper it cuts the geometry and the more is the noise level. Fig. 10 shows the test case geometries along with the enlarged cross-sectional area of the strut for the cases of no cutting element (Fig. 10a and e), cutting element edge length 3 voxels (Fig. 10b and f), 5 voxels (Fig. 10c and g) and 7 voxels (Fig. 10d and h).

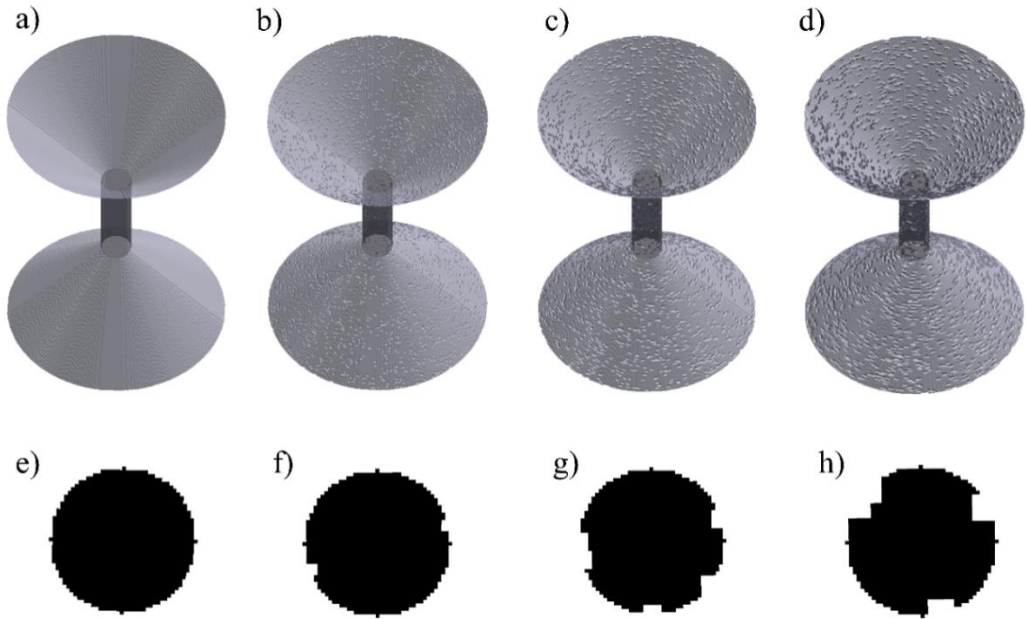

**Fig.10.** *Test case geometry with a) no cutting element, cutting element with edge length b) 3 voxels, c) 5voxels and d) 7 voxels; cross section of strut with e) no cutting element, cutting element with edge length f) 3 voxels, g) 5 voxels and h) 7 voxels.*

It can be seen in Figs. 10e-h that the depth of the cuts increase as the edge length of the cutting element is increased. Skeletonization algorithm is applied to all the four cases described in Fig. 10. The geometry of the test case is shown in light grey color. The resulting skeleton in shown as black color voxels points in the first column of Fig. 11a-d. It can be seen that the generated skeleton is highly sensitive to the surface noise. The complexity of the skeleton along with the number of the skeletal branches increases as the surface noise is increased from Fig. 11a to 11d. After pruning the free branches in all the four cases, the number of spurious branches reduce considerably. Since the task is to correctly identify the strut region, only those skeletal branches are important that define the strut region and everything else needs to be pruned. This is done by applying the pruning strategy of minimum cross-sectional area. Since, there is only strut in the test case whose cross-sectional area is already known from the diameter, only those skeletal branches are kept whose minimum cross-sectional area is closer to that of the strut. The threshold value for $\lambda_{minA}$ is chosen from the cross-sectional area of the strut without any noise (Fig. 11a). All other skeletal branches like the ones lying in the conical regions have higher cross-sectional areas. The results of pruning are shown in the third column in Fig. 11a-d. It can be seen that the skeletal branches that represent the strut region are correctly captured. Lastly, the strut volume is identified for all the cases as shown in the last column of Fig. 11a-d.

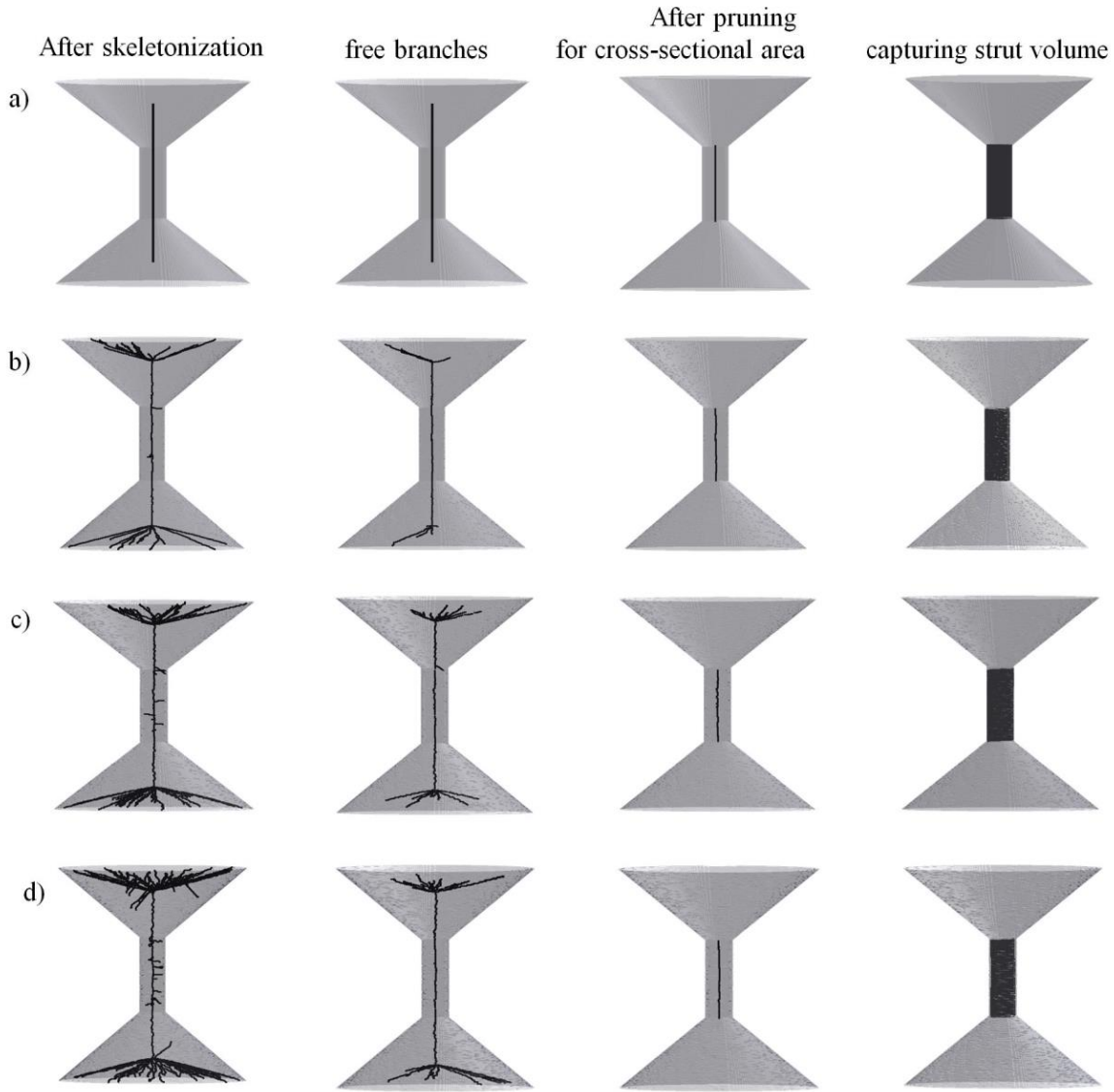

**Fig.11.** *Results of skeletonization and pruning strategies applied to test case geometry with a) no cutting element; cutting element with edge length: b) 3 voxels, c) 5 voxels and d) 7 voxels.*

### 3.1 Comparison with thickness-based thresholding

In the basic skeletonization algorithm, the thickness of the geometry is obtained from the voxel value of the distance transform of the image at the location of the skeletal points. However, identifying the struts on the basis of thickness is not possible as there are many skeletal branches in the conical regions that have the same thickness as that of the strut. Fig. 12a-d shows the struts segmented in test case geometries with different noise levels using threshold of thickness, instead of cross-sectional area. Any skeletal branch with minimum thickness less than 20.5 (radius of the strut) is segmented. It can be seen that thickness-based thresholding leads to unrealistic segmentation in the presence of surface noise as it shows slender features or struts in conical regions as well. Utilizing cross-sectional area instead of thickness as the pruning parameter helps in resolving this issue (last column Fig. 11).

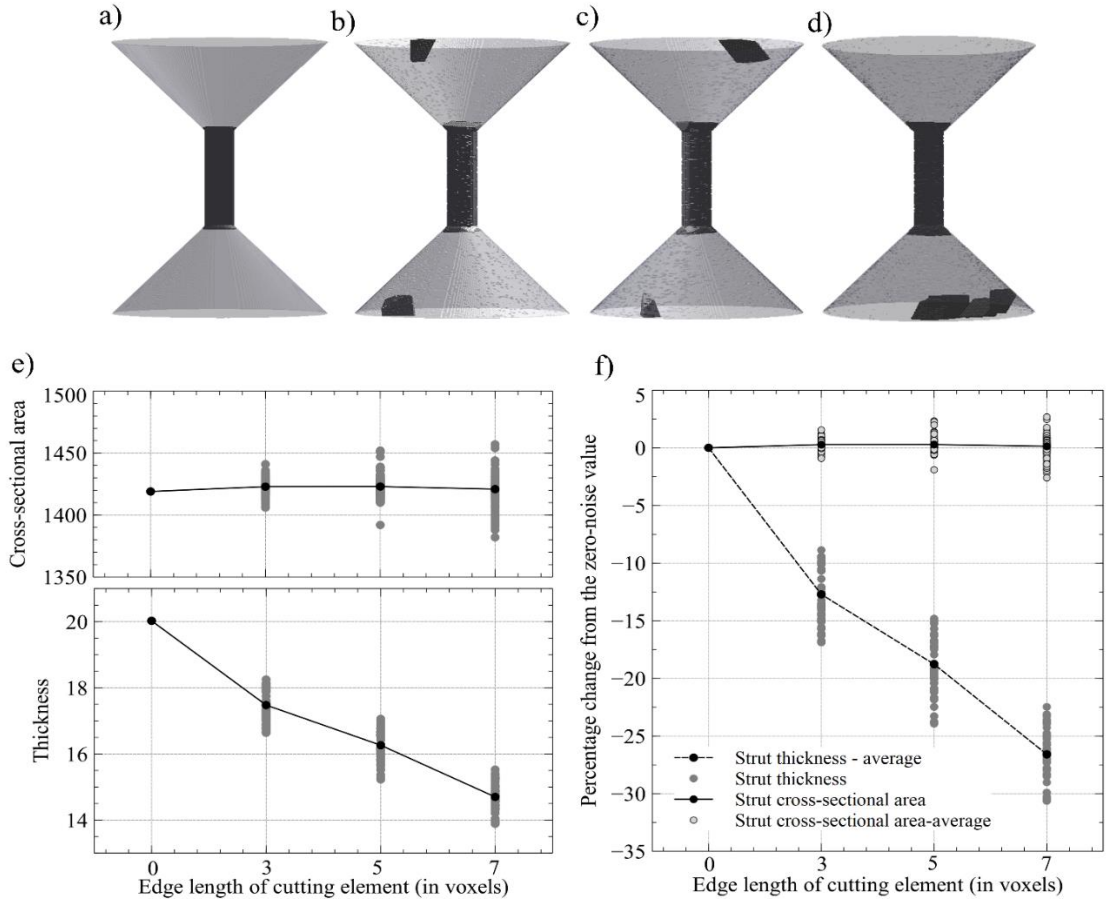

**Fig.12**. *Struts segmentation with thickness-based threshold in test case geometries with a) no cutting element; cutting element with edge length: b) 3 voxels, c) 5 voxels and d) 7 voxels; e) absolute values and f) relative change in thickness and cross-sectional area of the strut along its length for different values of edge length of the cutting element (noise levels).*

Another problem with thickness-based measurement is quantifying the segmented strut. To study this, the strut thickness as well as cross-sectional area is calculated at all the points along the skeletal branch that represents the strut. In an idealized strut without any surface noise, the thickness and cross-sectional area should be constant along the length of the strut. But due to surface noise, there is variation in this data. Fig. 12e shows thickness and cross-sectional area of the strut at different points along its length for test case geometries with different noise levels. The relative change in their values from the zero-noise case (0 edge length of cutting element) as the noise is increased in shown in Fig. 12f. It can be seen that as the noise is increased, the accuracy of the thickness prediction reduces drastically. But the cross-sectional area is not that sensitive to noise as the average value remains unchanged with less scatter compared to the thickness prediction. This shows that the cross-sectional area is a more robust parameter in not only identifying the strut regions in a geometry but also in correctly calculating the geometric properties of the strut.

## 4. Numerical results: Implementation on ceramic foam microstructure

The skeletonization algorithm is implemented in MATLAB [35] on the cubic domain of the foam microstructure shown in Fig. 2b. A small part of this domain is shown in light grey color in Fig. 13a along with the skeleton in black color.

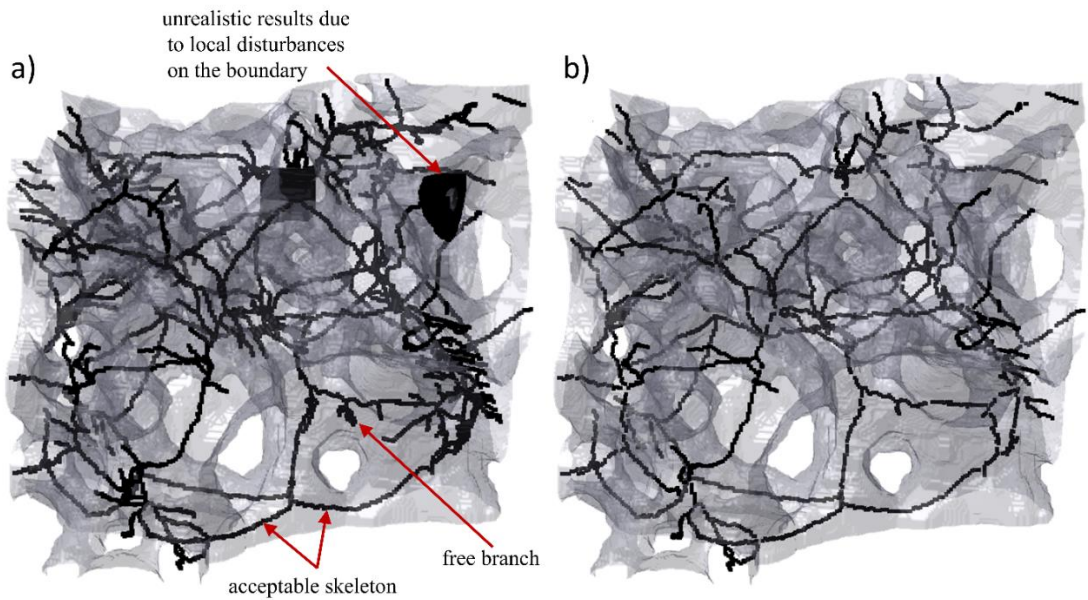

**Fig.13.** *A small part of the microstructure showing foam structure (grey color) and skeleton (black color).*

It can be seen that in some regions, the skeleton captures the geometry very well. But in some regions, there are unrealistic skeletal branches. One can also see some spurious free branches that do not represent any geometrical feature. Such spurious skeletal branches are created because of perturbations or noise on the surface of the foam structure which in turn arise from limitations in the resolution of the digital image. After pruning the free branches, the resulting skeleton of the same part of the microstructure is shown in Fig. 13b. It can be seen that a lot of spurious free branches have been removed by this pruning strategy. Note that the skeletal branches look disconnected because the junction points have been removed from the skeleton.

### 4.1 Pruning for geometric significance measures

Presence of surface noise leads to a lot of artificial skeletal branches especially in the junction regions of the microstructure. The geometry of these skeletal branches gives quantitative information that can be used to identify and prune these branches.

### 4.1.1 Length of the skeletal branch

At the junction region of an idealized microstructure, the skeletal branches of different struts would intersect to form one junction point. In a real microstructure, these branches intersect at an offset which creates small branches which do not represent any strut (shown in red color in Fig. 14 insert). These branches are removed by visually observing their length and apply a threshold value to prune them. Some of the examples of such artefact skeletal branches in the microstructure are shown in red color in Fig. 14. The threshold value is taken as $\lambda_{length} = 3$.

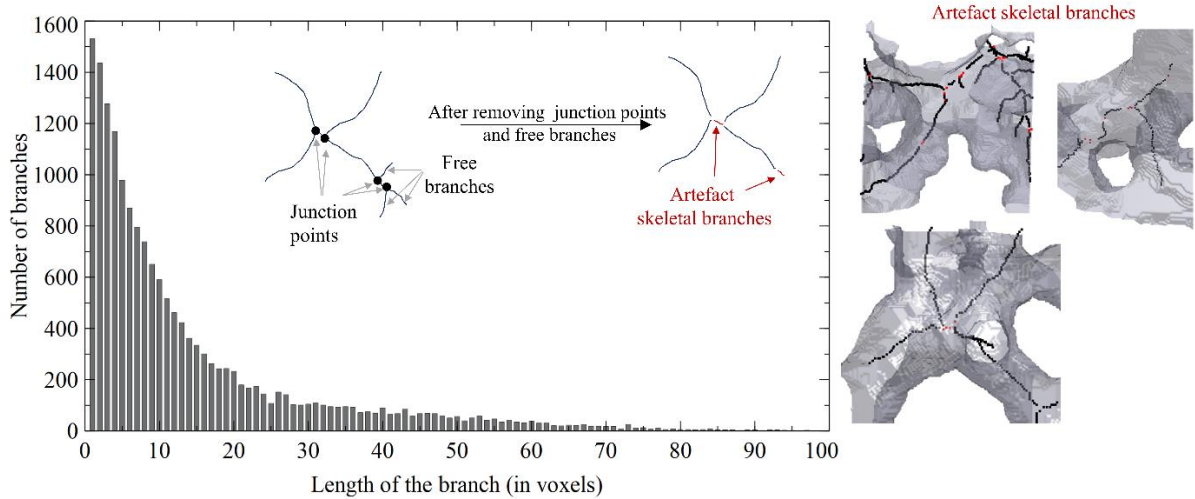

**Fig. 14.** *Length distribution of the skeletal branches in the cubic domain and examples of artefacts.*

.

### 4.1.2 Loop ratio of the skeletal branch

At some junction regions, there are narrow cavities which create a loop-like skeletal branches that do not represent the microstructure geometry. Examples of such cavities and artefact skeletal branches are shown in Fig. 15. Such artefacts are not a common occurrence in the microstructure and hence are outliers in the distribution of the loop ratio of the skeletal branches in the microstructure (left side of Fig. 15). They can be easily identified from the distribution and observed individually to decide the threshold value. It can be seen that skeletal branches with loop ratio lower than 0.6 are outliers in the distribution. In the studied microstructure, the threshold value is chosen as $\lambda_{LR}$=0.2.

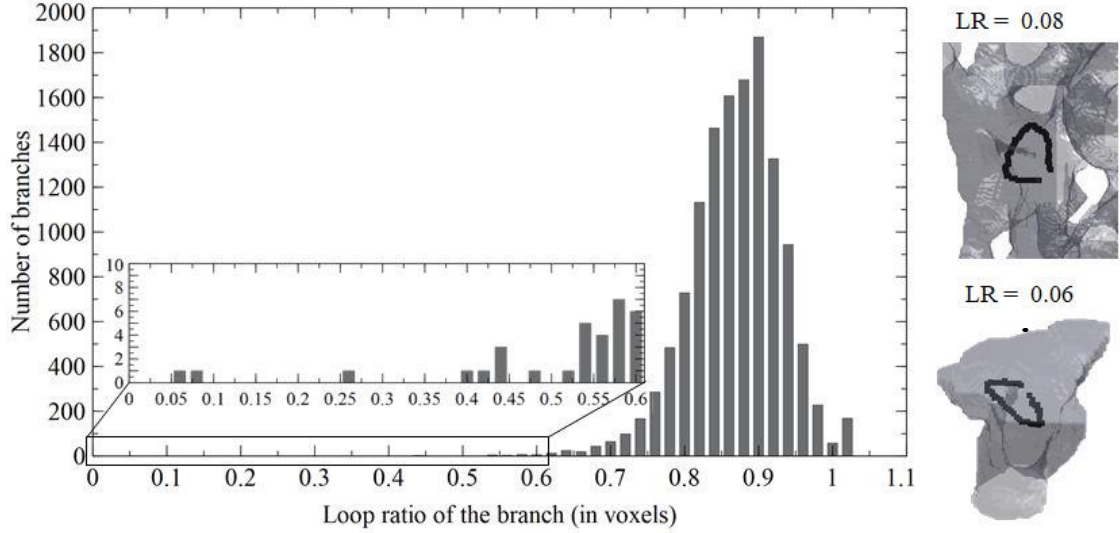

**Fig.15.** *Distribution of Loop Ratio of the skeletal branches in the cubic domain and examples of artefacts.*

### 4.1.4  Eccentricity of the cross-section of the strut

Some skeletal branches in the junction regions have very low thickness at a point close to the surface. But their cross-sectional area at that point is very high. This result is high eccentricity as defined in Eq.25. Examples of such skeletal branches and the points with maximum eccentricity are shown in Fig. 16. Such artefacts are also not a common occurrence in the microstructure and hence are outliers in the distribution of the eccentricity of the skeletal branches in the microstructure (left side of Fig. 16). They can be easily identified from the distribution and observed individually to decide the threshold value. It can be seen that skeletal branches with eccentricity higher than 10 are outliers in the distribution. In the studied microstructure, the threshold value is chosen as $\lambda_{eccen}=10$.

### 4.1.5  Aspect ratio of the strut

Another way to detect artificial skeletal branches in the junction region is to reconstruct the strut volume from these branches and calculate their aspect ratio. These struts have very low aspect ratios as shown in some examples in Fig. 17. Such branches are pruned by calculating their aspect ratios and then deciding a threshold value by visual observation. In the present microstructure, this value is taken as $\lambda_{AR} = 0.1$.

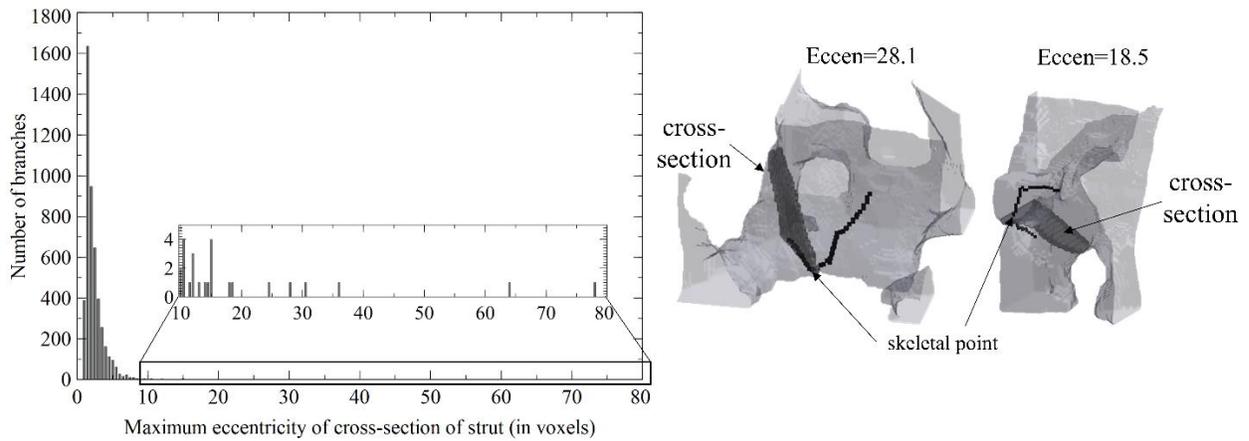

**Fig.16.** *Distribution of maximum eccentricity of the cross-sectional area of the skeletal branches in the cubic domain and examples of artefacts.*

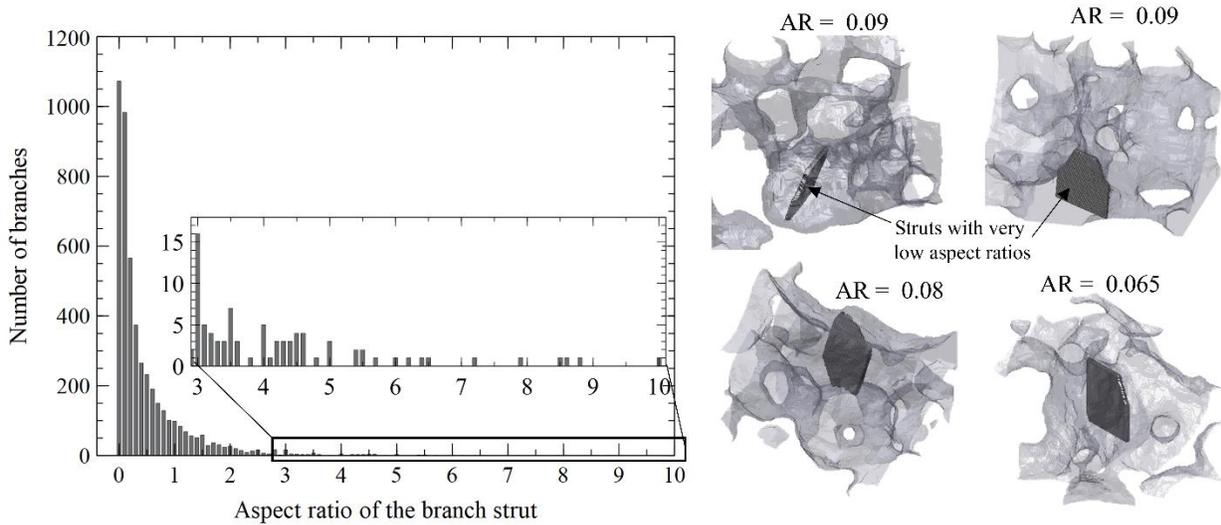

**Fig.17.** *Distribution of aspect ratio of the struts within the cubic domain and examples of artefacts.*

The main contribution of this paper is the pruning strategy based on cross-sectional area of the strut. However, because of the variety of the artificial skeletal branches existing in the microstructure, other pruning strategies explained before are required so as to clean the design space of the skeletal branches.

### 4.1.3 Cross-sectional area of the strut

The remaining skeletal branches are then pruned on the basis of their cross-sectional area. The objective here is not just to prune artificial branches but also to prune any branch whose cross-sectional area is outside the region of interest. The distribution of cross-sectional area of the struts in the microstructure as shown in Fig. 18 can aide is deciding the threshold value. In a porous microstructure, there will be large number of struts and hence the number of skeletal branches whose cross-sectional area correspond to that of the struts would be large. This number would drastically reduce for higher cross-sectional area that do not represent struts but some junction regions. Fig. 18 shows that the first bar with 2000 cross-sectional area has the highest number which decreases drastically for the next bar and so on. Visually observing the struts in these bars showed that the threshold value of 4000 captures all the struts in the microstructure. This threshold value is an input to the algorithm and can also come from various other sources depending on the application in question. Generally, the range of information of such slender regions is known a-priori from 2D microstructure images or from physics-based simulations.

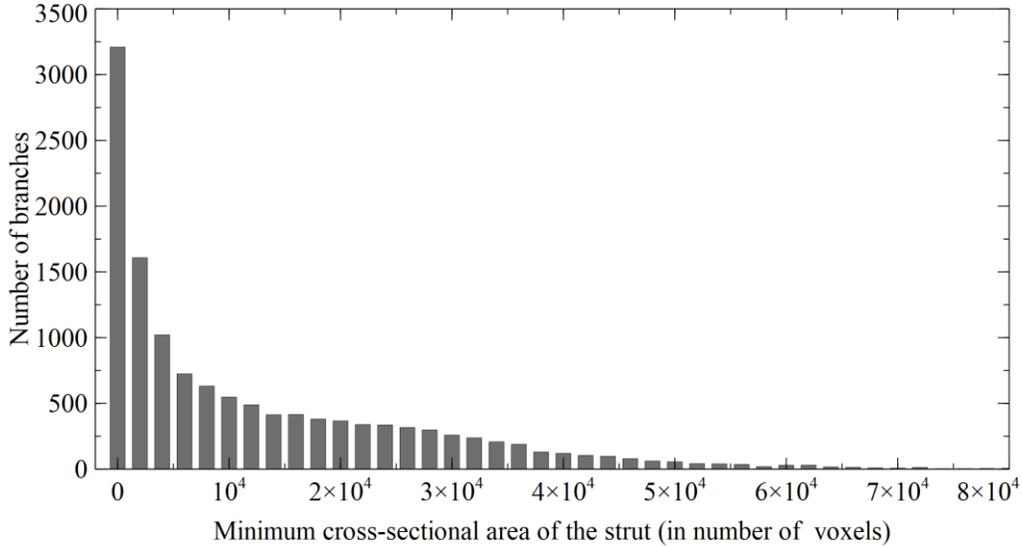

**Fig.18.** *Distribution of minimum cross-sectional area of the skeletal branches within the cubic domain.*

Note that the order in which the geometric pruning strategies are applied is not important as they are not dependent on each other. The general idea is to decide the order based on the computational effort required with the ones requiring less effort applied first.

### 4.2 Effectiveness of pruning strategies

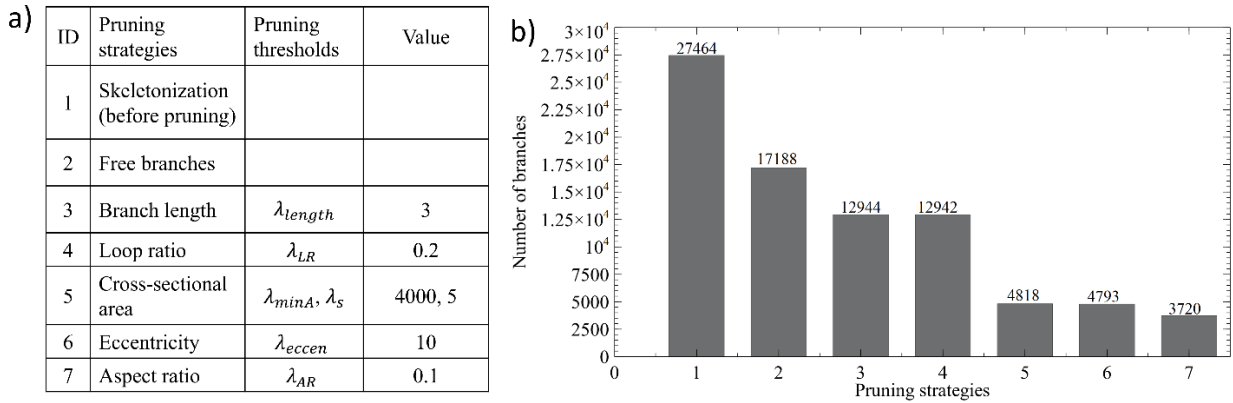

**Fig.19.** *a) Pruning strategies with IDs and values of the pruning parameters; c) Number of skeletal branches removed by every pruning strategy.*

Each pruning strategy (Fig.19a) removed a certain number of skeletal branches. Fig. 19b shows the number of skeletal branches remaining after every pruning strategy. The X-axis of Fig. 19b corresponds to IDs of pruning strategies. It can be seen that there is a significant reduction in number of skeletal branches after pruning the free branches, pruning for branch length, cross-sectional area and aspect ratio. Note that only 3720, i.e. about 13.5 % of the total skeletal branches (27464) are remaining after applying all the pruning strategies. This number is obviously dependent on the problem at hand especially while deciding the threshold value of minimum cross-sectional area, $\lambda_{minA}$.

Finally, some examples of strut volumes segmented by the algorithm have been shown in Fig. 20. It can be seen that these slender regions have arbitrary shapes yet the algorithm is capable of isolating them from the neighboring thick junction regions.

The algorithm can be utilized to study a variety of microstructures where the objective is to isolate slender regions within a phase. In order to demonstrate this, the algorithm is implemented on different microstructures as shown in section A1 in the supplementary document. It shows two numerically reconstructed microstructures (explained in detail in [31]) of ceramic foam of different volume fractions. The cross-sectional area and aspect ratio distribution of the struts in these microstructures is different from the one presented in the main paper because of the difference in the volume fraction. Yet, the algorithm is able to segment the struts effectively. Next, an architected foam having IWP-type triply periodic minimal surface [36] is studied which has a gradient volume fraction. The struts are thinnest on top right corner and thickest on bottom right corner. As the threshold value of cross-sectional area is increased, the algorithm segments the corresponding struts effectively.

## 5. Conclusions

The article describes a skeletonization based algorithm to segment slender regions in a 3D microstructure. Its particular use is in isolating slender regions which are connected to thick regions within the same phase of the microstructure. To overcome the sensitivity of the skeletonization algorithm to surface noise, novel pruning strategies have been devised. The strategy of pruning free branches removes the hairy skeletal branches that do not represent the actual geometry of the microstructure. The geometric pruning strategies of length, loop ratio, eccentricity and aspect ratio remove any spurious branches or artefacts that do not represent the actual slender regions but are the result of insufficient image resolution. The pruning strategy based on cross-sectional area captures the skeletal branches that represent the slender regions in the image depending on their minimum cross-sectional area. It also acts as a regularization to the skeletonization algorithm. The cross-sectional area is not as sensitive to surface noise as the thickness on which the skeletonization algorithm is based. Results of the test case in Fig. 11 show that this pruning strategy is successful in segmenting strut (slender region) irrespective of the noise level on the surface of the geometry. Figs. 12e-f show that the thickness of the slender region obtained from distance transform changes drastically with surface noise. On the contrary, the surface noise has negligible effect on the cross-sectional area. The implementation of the algorithm on the microstructure of ceramic foam shows that it is capable of segmenting struts that have arbitrary shapes and cross-sections. It is capable of segmenting slender regions that have a wide range of distribution in their geometric properties of length, cross-sectional area and aspect ratio. The use of threshold values for different pruning strategies makes the algorithm adaptable to different types of structures not only limited to material microstructures but also to any type of 3D image. It can provide valuable insights to any physical problems in material science where the slender regions of the material microstructure play a critical role.

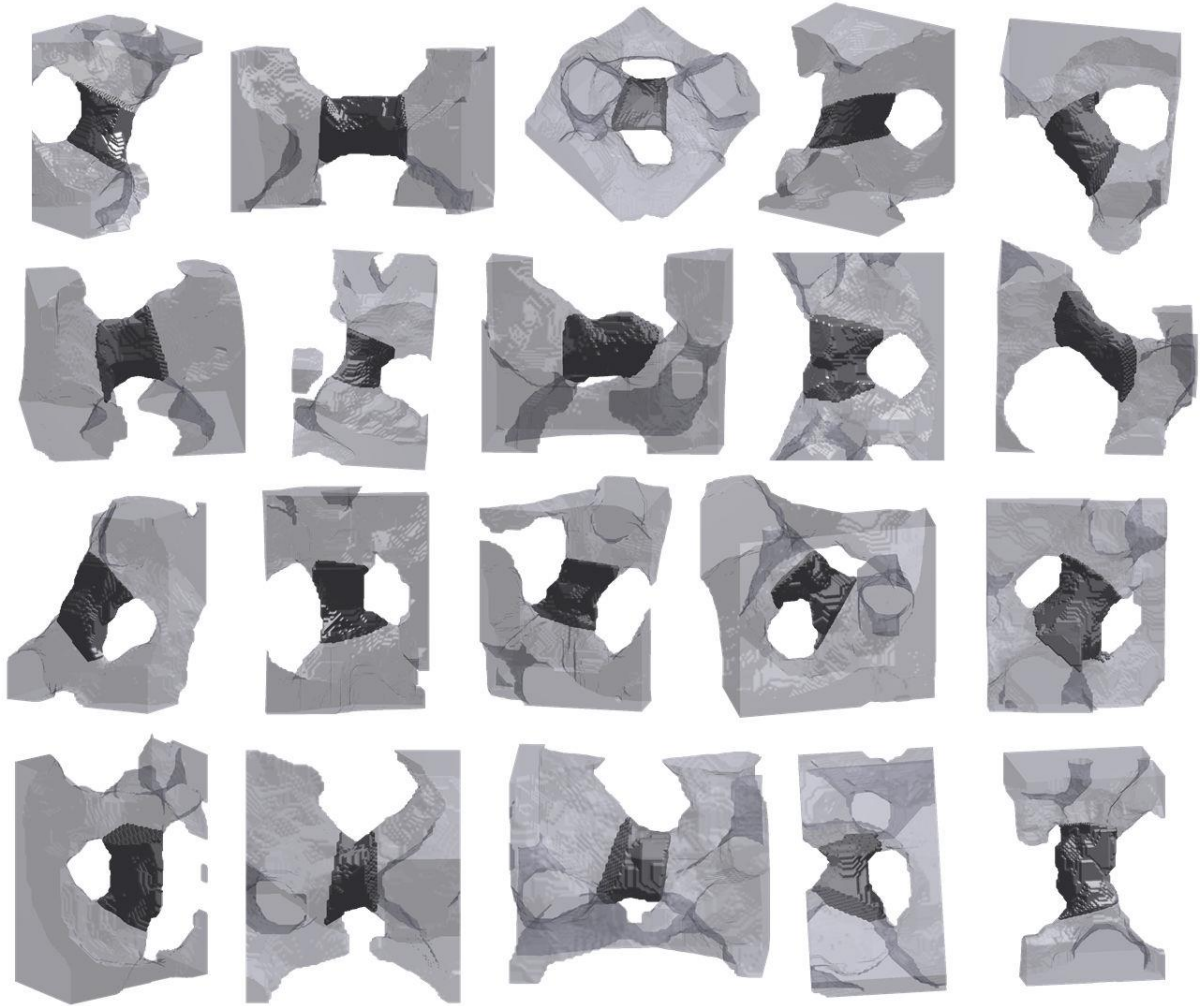

**Fig.20.** *Examples of strut volumes segmented by the developed algorithm.*

**Appendix**

Section A1 in the supplementary document shows the implementation of the algorithm on different microstructures. Section A2 describes the algorithms implemented in this work. The data file of the ceramic foam microstructure and the MATLAB codes can be found in GitHub repository: https://github.com/Vinit-Deshpande/Segment-slender-regions.git

**Acknowledgements**

The financial support of the University of Applied Sciences Darmstadt is gratefully acknowledged. This research was supported by the Hessian Ministry of Higher Education, Research, Science and the Arts, Germany within the Framework of the "Programm zum Aufbau eines akademischen Mittelbaus an hessischen Hochschulen". This research was also funded by the Federal Ministry for Economic Affairs and Energy based on a decision of the German Bundestag.

**Declaration of competing interest**

The authors declare that they have no known competing financial interests or personal relationships that could have appeared to influence the work reported in this paper.